\documentclass[11pt]{article}
\usepackage{moriond,epsfig}
\usepackage[latin1]{inputenc}
\usepackage[OT1]{fontenc}

\def\be{\begin{equation}}
\def\ee{\end{equation}}
\def\bea{\begin{eqnarray}}
\def\eea{\end{eqnarray}}

\begin{document}
\vspace*{4cm}
\title{ON COSMOLOGICAL OBSERVABLES IN A SWISS-CHEESE UNIVERSE\footnote{Contribution to the proceedings of the 43rd Rencontres de Moriond, Cosmology Session, La Thuile, Italy, March 15-22, 2008.
Based on\cite{Marra:2007pm,Marra:2007gc}, where more details and references can be found.}}

\author{VALERIO MARRA}

\address{Dipartimento di Fisica ``G.\ Galilei'', Universit\`{a} di Padova; INFN Sezione di Padova \\
Via Marzolo 8, Padova I-35131, Italy. \  E-Mail: valerio.marra@pd.infn.it}

\maketitle\abstracts{
By means of a toy Swiss-cheese cosmological model we will discuss how to set up and carry out in a physically meaningful way the idea of back-reaction, according to which dark energy could be an effective source. We will follow two distinct approaches. One is focused on how cosmological observables are affected by inhomogeneities, while the other is focused on a theoretical description of the inhomogeneous universe by means of a mean-field description.
}

\section{A point of view}
The ``safe'' consequence of the success of the concordance model is that the
isotropic and homogeneous $\Lambda$CDM model is a good  {\it observational}
fit to the real inhomogeneous universe. And this is, in some sense, a
verification of the cosmological principle:  the inhomogeneous universe can be
described by means of an isotropic and homogeneous solution.\\
However, this does not imply that a primary source of dark energy  exists, but only that it
exists as far as the observational fit is concerned. For example, it is not
straightforward that the universe is accelerating. If dark energy does not
exist at a fundamental level, its presence in the concordance model would tell
us that the pure-matter inhomogeneous model has been renormalized, from the
observational point of view, into a homogeneous $\Lambda$CDM model\cite{Kolb:2005da,Rasanen:2003fy}.

There are, broadly speaking, two distinct approaches about smoothing out inhomogeneities\cite{Marra:2008sy}. One is focused on how cosmological observables are affected by inhomogeneities, while the other rests on a theoretical description of the inhomogeneous universe by means of a mean field.

We studied both approaches by means of a toy Swiss-cheese cosmological model\cite{Marra:2007pm,Marra:2007gc}, where the cheese is made of the EdS model and the holes are constructed from a Lema\^{\i}tre-Tolman-Bondi solution of Einstein's equations. LTB solutions are based on the assumption of spherical symmetry, the motion is radial and geodesic.
It has been indeed shown that the LTB solution can be used to fit the observed
$d_{L}(z)$  without the need of dark energy\cite{Alnes:2005rw,Alexander:2007xx,Apostolopoulos:2006eg}. To achieve this result, however, it is necessary to place the observer at the center of a rather big underdensity.\\
To overcome this fine-tuning problem we built a Swiss-cheese model with the observer in
the cheese looking through a series of holes.


\section{Swiss-cheese}

\begin{figure}[h!]
\begin{center}
$\begin{array}{c@{\hspace{15mm}}c}
\includegraphics[width=7.1 cm]{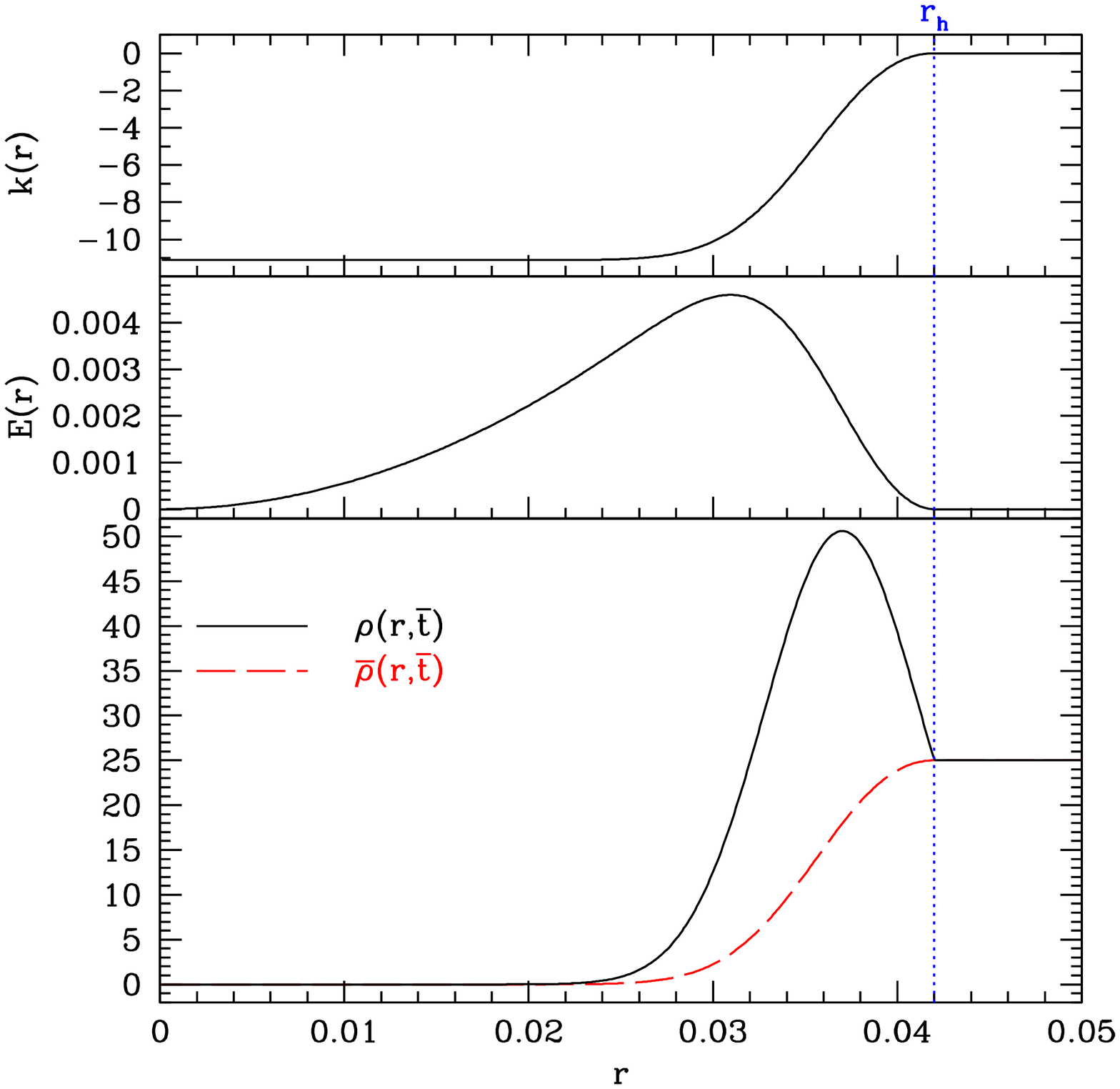}  &  
\includegraphics[width=7.1 cm]{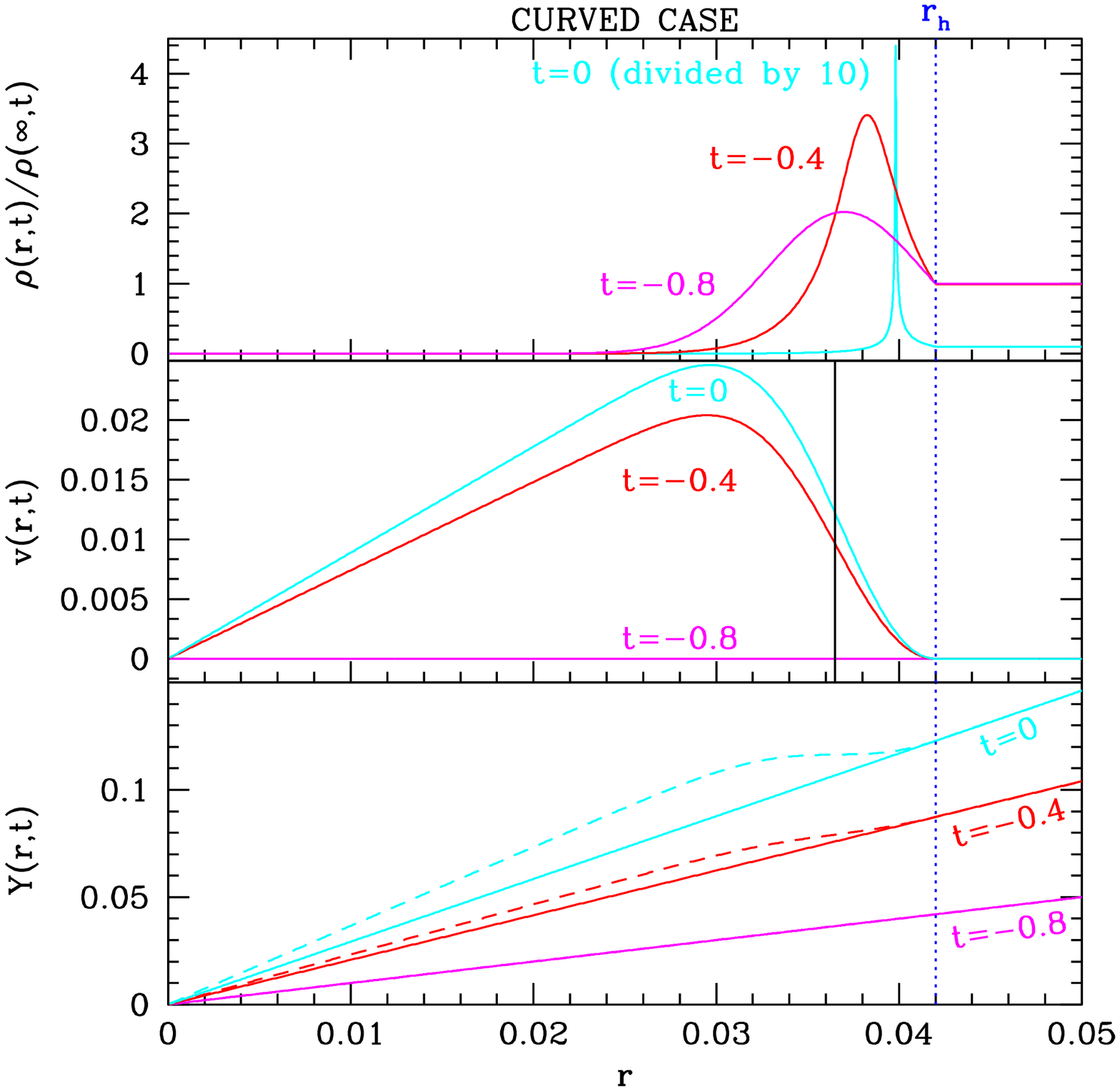}   \\
\end{array}$
\caption{\scriptsize{Left panel, bottom: the densities $\rho(r,\bar{t})$ (solid curve) and
$\bar{\rho}(r,\bar{t})$  (dashed curve). The matching to the FRW solution
is achieved  as one can see from the plot of $\bar{\rho}(r,\bar{t})$.
Top: curvature $k(r)$ and $E(r)$ necessary for the 
initial conditions of no peculiar velocities.
On the right dynamics of the LTB solution. From the top to the bottom: energy density, peculiar velocities and scale factor.
Here, $\bar{t}=-0.8$ corresponds to $z\simeq 2$.
The hole ends at $r_{h}=0.042$ which corresponds to $350$Mpc.}}
\label{sc}
\end{center}
\end{figure}

In this section we will briefly describe our Swiss-cheese. In Fig.~\ref{sc} you can see, on the left panel, the density and curvature which characterize our LTB model: there is a void at the center of the holes which is dominated by negative curvature. Then there is a compensating overdensity and finally both the curvature and the (averaged) density match their EdS value at the border of the hole. The matching is necessary in order to have an exactly solvable Swiss-cheese model.
The Swiss cheese we have in mind is a lattice of identical holes of radius $r_{h}\simeq 350$Mpc.

On the right panel of Fig.~\ref{sc} the dynamics is shown. The void is expanding faster than the cheese causing the formation of a thin shell and of a large void.

\section{Luminosity-distance--redshift relation}

\begin{figure}[h!]
\begin{center}
$\begin{array}{c@{\hspace{15mm}}c}
\includegraphics[width=7 cm]{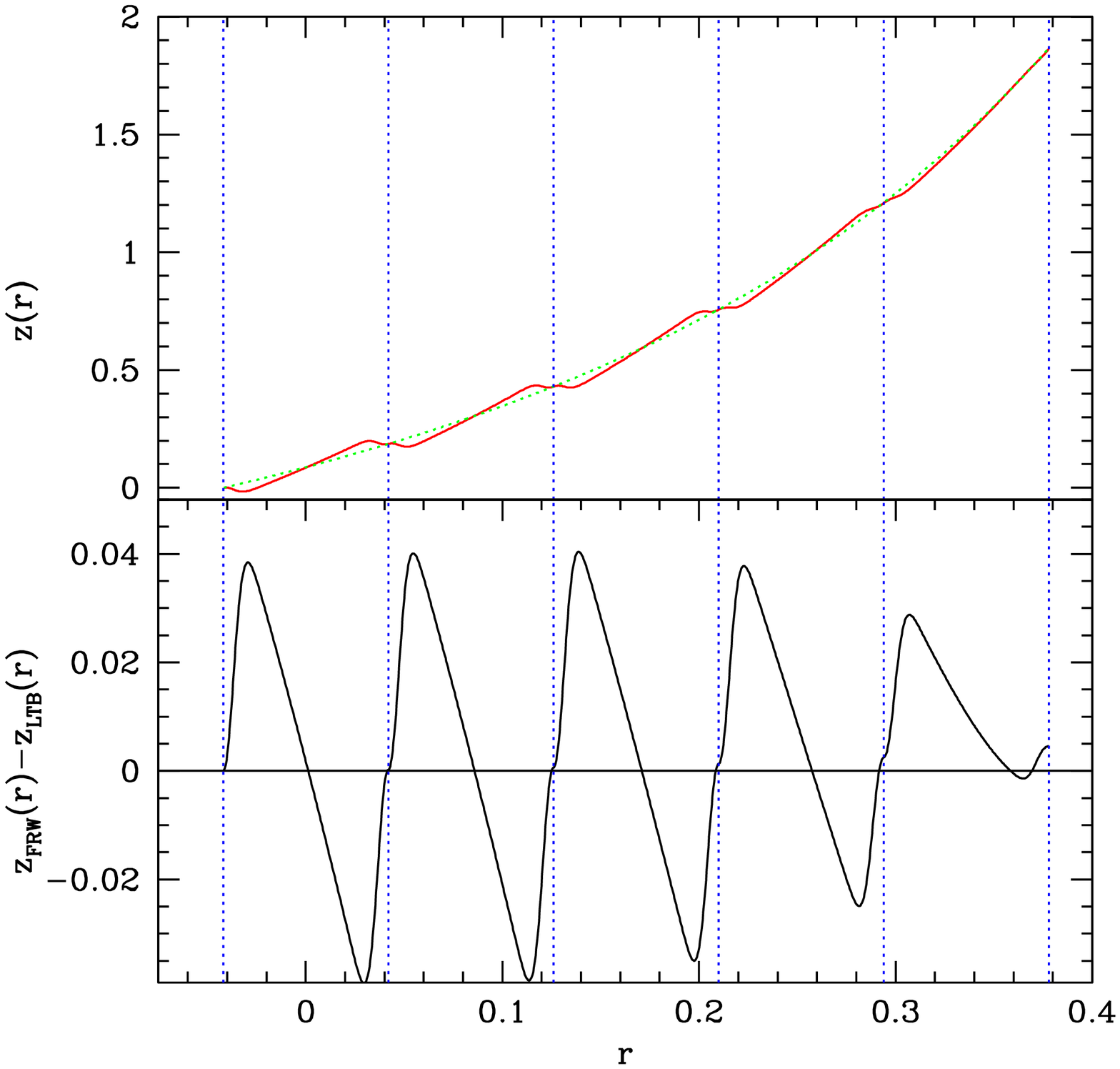}  &  
\includegraphics[width=7 cm]{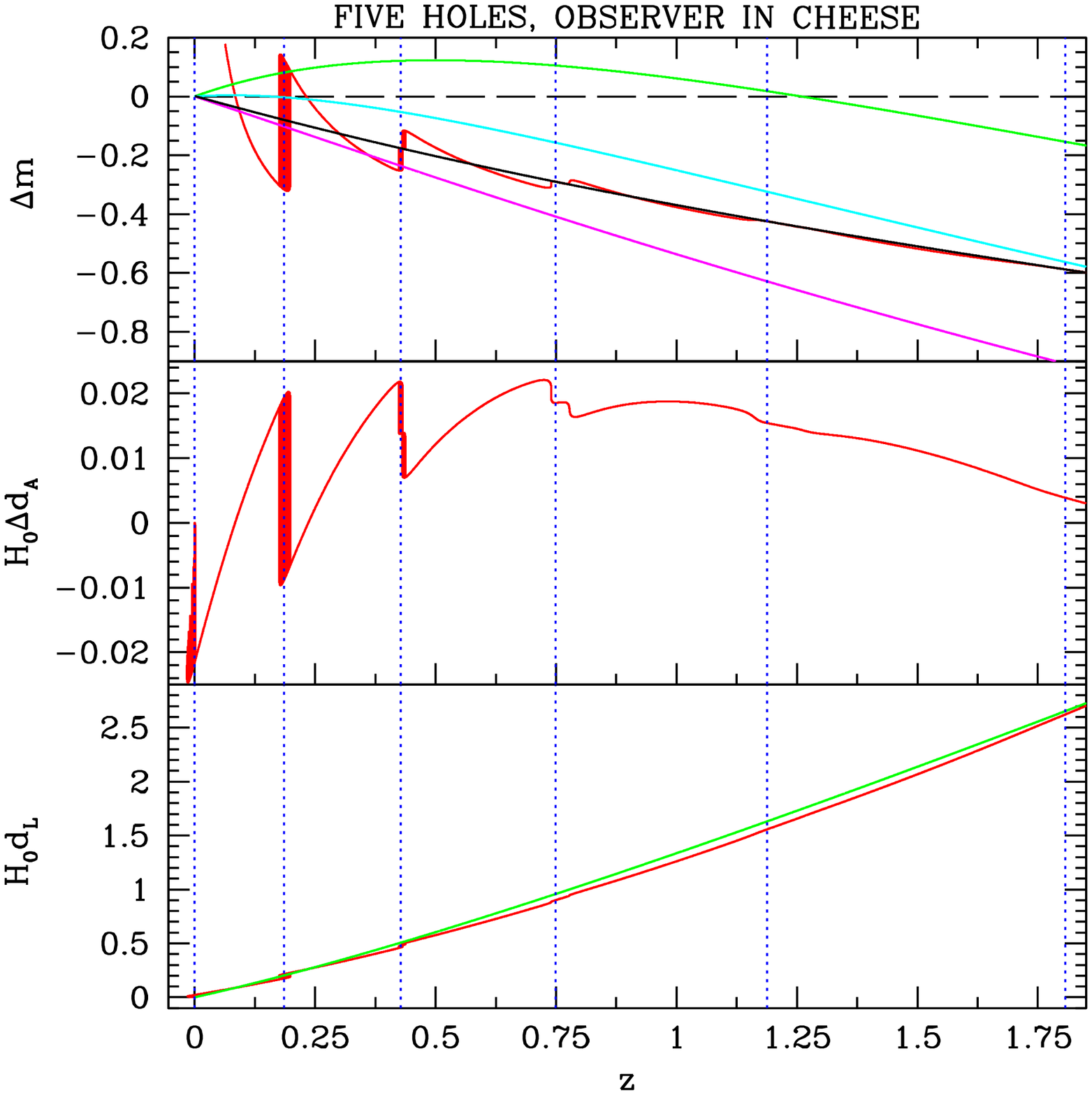}   \\
\end{array}$
\end{center}
\caption{\scriptsize{On the left, redshift histories for a photon that travels from one side of the five-hole chain to the other where the 
observer will detect it at present time. The ``regular'' curve is for the EdS
model. The vertical lines mark the edges of the holes.
On the right, the bottom shows the luminosity distance $d_L(z)$ for the Swiss-cheese
(jagged curve) and the $\Lambda$CDM solution with $\Omega_{M}=0.6$ and
$\Omega_{DE}=0.4$ (regular curve).  In the middle is the change in
the angular diameter distance, $\Delta d_A(z)$, compared to the same $\Lambda$CDM
model. The top panel shows the
distance modulus in various cosmological models. The jagged line is for the
Swiss cheese. The regular curves, from top to bottom, are a $\Lambda$CDM
model with  $\Omega_{M}=0.3$ and $\Omega_{DE}=0.7$, a $\Lambda$CDM model with  
$\Omega_{M}=0.6$ and $\Omega_{DE}=0.4$, the best smooth fit to the LTB model,
and the EdS model.}}
\label{dlz}
\end{figure}

In this section we will be interested in how cosmological observables are affected by inhomogeneities.
In our set up the observer is in the cheese and is looking through a series of five holes.
We will focus on the luminosity-distance--redshift relation $d_{L}=(1+z)^{2} \, d_{A}$: we will show in Fig.~\ref{dlz} on the left how redshift changes, while we will show on the right the results for $\Delta m$.

As you can see from Fig.~\ref{dlz} we found that redshift effects are suppressed because of
a compensation effect acting on the scale of half a hole, due to spherical
symmetry and the matching performed. However, we found interesting effects in the calculation of the angular
distance: the evolution of the inhomogeneities bends the photon path compared
to the EdS case. Therefore, inhomogeneities will be able (at least partly) to
mimic the effects of dark energy.

\section{Mean-field description}
\begin{figure}[h!]
\begin{center}
$\begin{array}{c@{\hspace{15 mm}}c}
\includegraphics[width=7 cm]{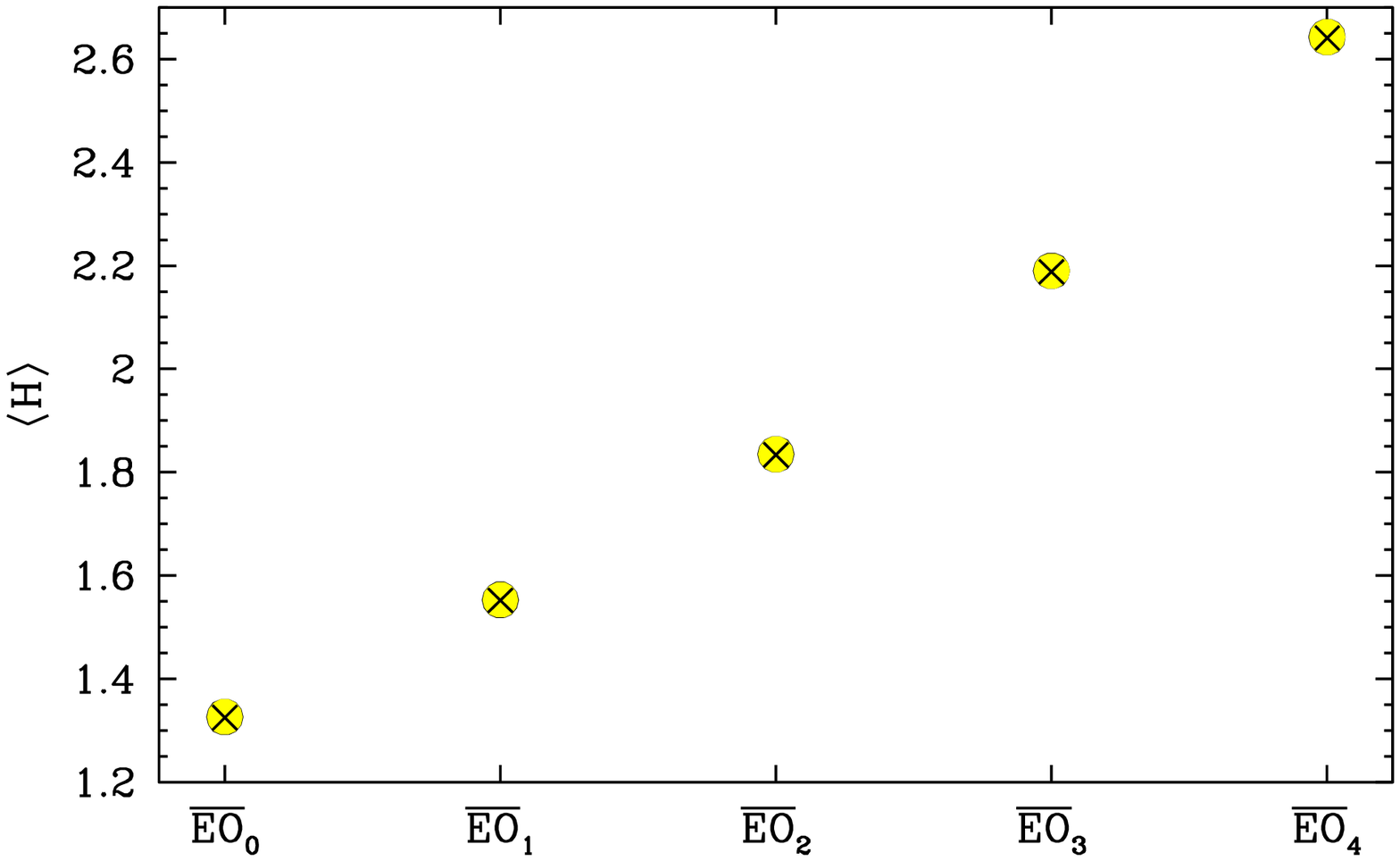}  &  
\includegraphics[width=7 cm]{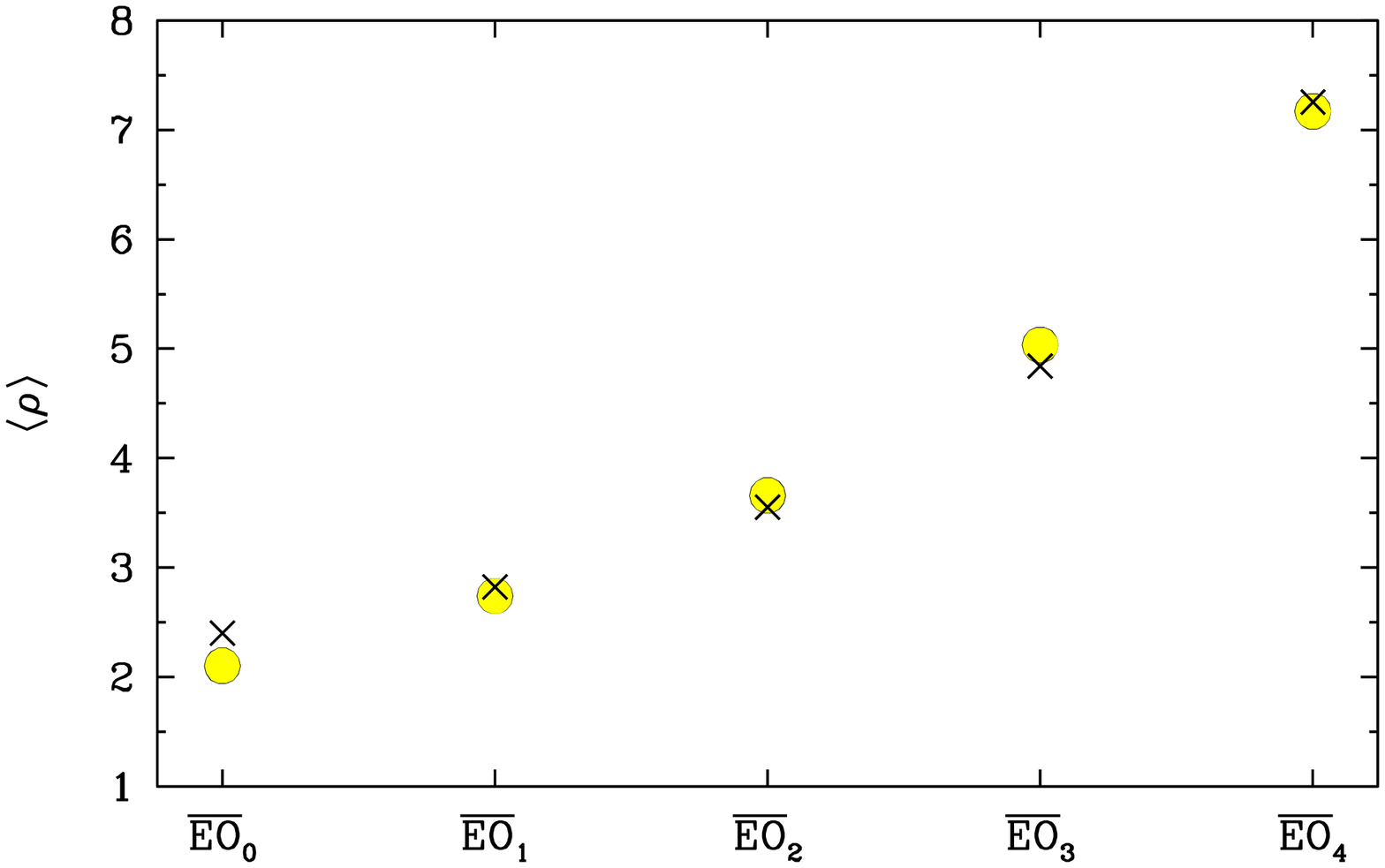}   \\
\end{array}$
\end{center}
\caption{\scriptsize{If we consider the expansion rate (left panel) we will find a phenomenological model with $w\simeq 0$, while if we consider the density (right panel) we will find a phenomenological model with $w_{0}=-1.95$ and $w_{a}=4.28$.}}
\label{fit}
\end{figure}

In this section we will focus on the second possible approach which aims at finding a mean-filed description of the Swiss-cheese universe.
To this end we will study the fitting approach\cite{ellis-f}, in particular we will fit by means of light-cone averages a quintessence-like phenomenological model to our Swiss-cheese. We will assume that the dark-energy component in the phenomenological model has an equation of state $w=w_0+w_a \; z/ (1+z)$.
You can see the results in Fig.~\ref{fit}. If we consider the expansion (left panel) we will simply find the EdS model as the best-fit phenomenological model, or in other words, we did not find any effects and this is due to the compensating effect of spherical symmetry.
The situation is, however, different if we consider density (right panel): the photon is spending more and
more time in the (large) voids than in the (thin) high density structures.

Summarizing, in a Swiss-cheese model with spherically symmetric holes photon redshift is not affected by inhomogeneities. A photon spending most of its time in voids should have, however, a different redshift history.
We have noticed, however, that density is not particularly sensitive to spherical symmetry and therefore we can argue that a Swiss-cheese made of spherically symmetric holes and one made of non-spherically symmetric holes will share the same light-cone averaged density. This step is pictured below.

\begin{figure}[h!]
\begin{center}
\includegraphics[width=8.0 cm]{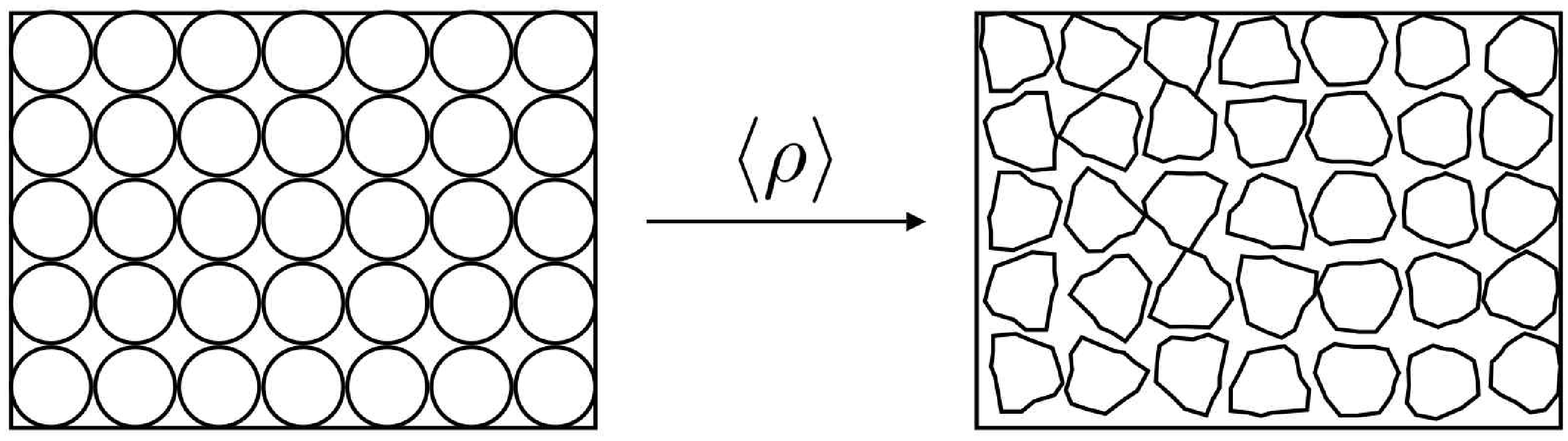}
\end{center}
\end{figure}

\clearpage

We will then get the expansion (and so the redshift history) from the density.
At the end, we will end up with the phenomenological best-fit model we have previously found which behaves similarly to the concordance model.

\section{Conclusions}

To make the point we will show in a flow chart what we have found by means of the two different approaches we have followed.
Regarding $d_{L}(z)$, we found no important effects from a change in the
redshift:  the effects on $d_L(z)$ all came from $d_{A}$ driven by the evolution
of the inhomogeneities.
Regarding light-cone averages, we found no important effects with respect 
to the expansion: this negative result is due to the compensation in the
previously discussed redshift and it is the same reason 
why we did not find redshift effects with $d_{L}(z)$.
We found important effects with respect to the density: however this is 
not due to the effects driving the change in $d_{A}$. The latter is due to 
structure evolution while the former to the presence of voids, so the 
two causes are not directly connected.

\begin{figure}[h!]
\begin{center}
\includegraphics[width=11 cm]{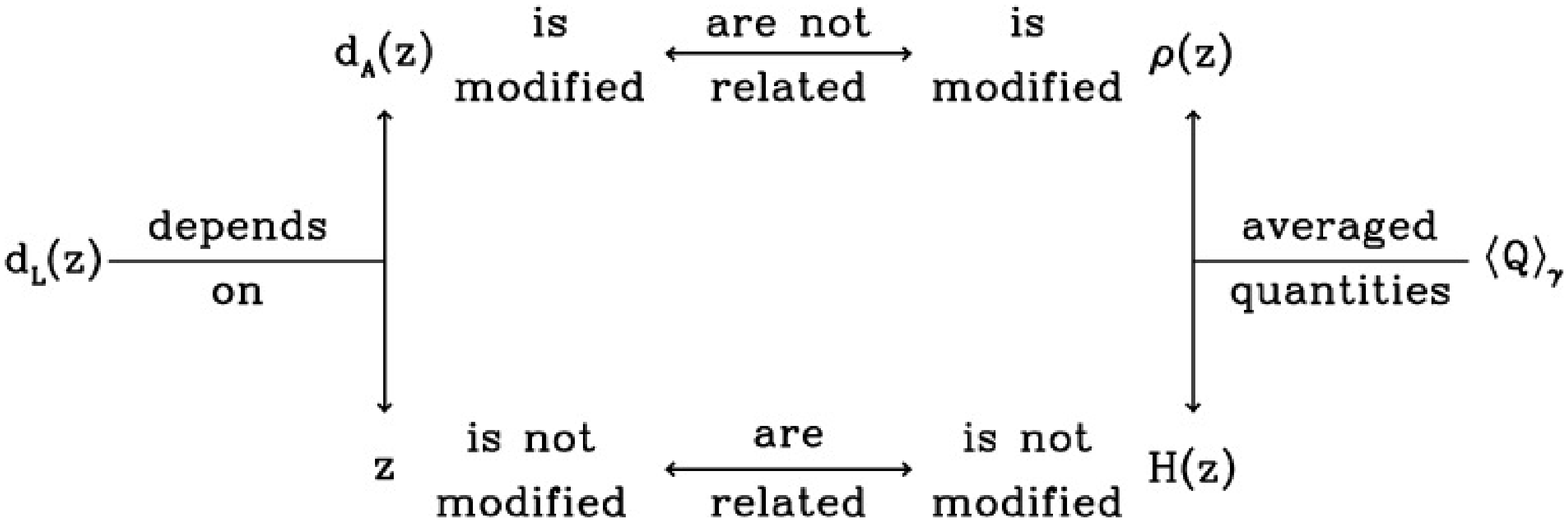}
\end{center}
\end{figure}

To conclude we would like to stress the importance of setting up and carrying out in a physically meaningful way the idea of the back-reaction or, in other words, of taking into account the smoothing of inhomogeneities. Different questions will indeed give different answers.

\section*{Acknowledgements}
I have used work done in collaboration with E.~W.~Kolb, S.~Matarrese and A.~Riotto.
V.M. acknowledges support from ``Fondazione Ing. Aldo Gini'' and ``Fondazione Angelo Della Riccia''.

\end{document}